\documentclass[runningheads]{llncs}
\usepackage{graphicx}
\usepackage[table,xcdraw,dvipsnames]{xcolor}
\usepackage{amsmath}
\usepackage{booktabs}
\usepackage{arydshln}
\usepackage{multirow}
\usepackage[normalem]{ulem}
\usepackage{pdflscape}
\usepackage{enumitem}
\usepackage[hidelinks]{hyperref}
\usepackage{pifont}
\usepackage{arydshln}

\newcommand{\cmark}{{\color{ForestGreen}\ding{51}}}%
\newcommand{\xmark}{{\color{BrickRed}\ding{55}}}%

\newcommand{\set}[1]{\mathcal{#1}}

\newcommand{\mantis}{\texttt{MANtIS}}
\newcommand{\msdialog}{\texttt{MSDialog}}
\newcommand{\ubuntu}{\texttt{UDC\textsubscript{DSTC8}}}

\newcommand{\bm}{BM25}
\newcommand{\rmthree}{RM3}
\newcommand{\resptocontext}{resp2ctxt}
\newcommand{\resptocontextlu}{resp2ctxt$_{lu}$}

\newcommand{\usep}{[U]}
\newcommand{\tsep}{[T]}

\begin{document}
\title{Do the Findings of Document and Passage Retrieval Generalize to the Retrieval of Responses for Dialogues?}
\titlerunning{From Document and Passage Retrieval to Response Retrieval for Dialogues}

\author{Gustavo Penha  \and
Claudia Hauff}

\institute{TU Delft\\
\email{\{g.penha-1,c.hauff\}@tudelft.nl}}

%
%
%
\maketitle              
\begin{abstract}
A number of learned sparse and dense retrieval approaches have recently been proposed and proven effective in tasks such as passage retrieval and document retrieval. In this paper we analyze with a replicability study if the lessons learned generalize to the retrieval of responses for dialogues, an important task for the increasingly popular field of conversational search. Unlike passage and document retrieval where documents are usually longer than queries, in response ranking for dialogues the queries (dialogue contexts) are often longer than the documents (responses). Additionally, dialogues have a particular structure, i.e. multiple utterances by different users. With these differences in mind, we here evaluate how generalizable the following major findings from previous works are: \textbf{(F1)}~query expansion outperforms a no-expansion baseline; \textbf{(F2)}~document expansion outperforms a no-expansion baseline; \textbf{(F3)}~zero-shot dense retrieval underperforms sparse baselines; \textbf{(F4)}~dense retrieval outperforms sparse baselines; \textbf{(F5)}~hard negative sampling is better than random sampling for training dense models. Our experiments\footnote{\textcolor{RubineRed}{\url{https://github.com/Guzpenha/transformer\_rankers/tree/full\_rank\_retrieval\_dialogues}}.}---based on three different information-seeking dialogue datasets---reveal that four out of five findings (\textbf{F2}--\textbf{F5}) generalize to our domain.

\end{abstract}

\section{Introduction}\label{sec:intro}
\emph{Conversational search} is concerned with creating agents that satisfy an information need by means of a \emph{mixed-initiative} conversation through natural language interaction, rather than through the traditional search engine results page.  A popular approach to conversational search is retrieval-based~\cite{anand2020conversational}: given an ongoing conversation and a large corpus of historic conversations, retrieve the response that is best suited from the corpus~\cite{wu2017sequential,yang2018response,penha2019curriculum,yang2020iart,han-etal-2021-fine}. Due to the effectiveness of heavily pre-trained transformer-based language models such as BERT~\cite{devlin2018bert}, they have become the predominant approach for conversation response re-ranking~\cite{penha2019curriculum,whang2021response,zhang2021structural,gu2020speaker,whang2019effective}.


The most common evaluation procedure for conversation response re-ranking consists of re-ranking a limited set of $n$ candidate responses (including the ground-truth response(s)), followed by measuring the number of relevant responses found in the first $K$ positions---$Recall_n@K$~\cite{zhang2021advances}. Since the entire collection of available responses is typically way bigger\footnote{While for most benchmarks~\cite{zhang2021advances} we have only 10--100 candidates, a working system with the Reddit data from PolyAI \url{https://github.com/PolyAI-LDN/conversational-datasets} would need to retrieve from 3.7 billion responses.} than such a set of candidates, this setup is in fact a selection problem, where we have to choose the correct response out of a few options. This evaluation overlooks the first-stage retrieval step, which retrieves a set of $n$ responses to be re-ranked. If the first-stage model, e.g. BM25, fails to retrieve relevant responses, the entire pipeline fails.

Motivated by a lack of research on the first-stage retrieval step, we are interested in answering in our replicability study whether the considerable knowledge obtained on document and passage retrieval tasks generalizes to the dialogue domain. Unlike document and passage retrieval where the documents are generally longer than the queries, in response retrieval for dialogues the queries (dialogue contexts) tend to be longer than the documents (responses). A second important difference is the structure induced by the dialogue as seen in Table~\ref{table:intro}.

\setlength\dashlinedash{0.2pt}
\setlength\dashlinegap{1.5pt}
\setlength\arrayrulewidth{0.3pt}
\begin{table}[]
\centering
\label{table:intro}
\caption{Comparison between passage retrieval and response retrieval for dialogues. In \S\ref{section:method} we define the task of \emph{First-stage Retrieval for Dialogues}. Colors symbolize the {\color{BlueViolet}{information-seeker}} and {\color{purple}{information-provider}}. $p^+/r^+$ are the relevant passage/response.}
\begin{tabular}{@{}p{2cm}p{3.5cm}p{0.5cm}p{5.6cm}@{}}
\toprule
 & \textbf{Passage Retrieval} & & \textbf{First-stage Retrieval for Dialogues} \\ \midrule
\textbf{Input} & Query $q$ & & Dialogue context $\set{U} = \{u^1, u^2, ... , u^{\tau}\}$\\ \hdashline
\textbf{Example} &$q$: \textit{what is theraderm used for} & &
\begin{tabular}[c]{@{}p{5cm}@{}}{ $u^1$: \color{BlueViolet}\textit{I want a firewall that will protect me but more of that to monitor any connection in or out of my mac} [...]}\\$u^2$:  {\color{purple} \{url\} \textit{allows you to map joypad buttons to keyboard keys and [...]}} \\$u^3$: {\color{BlueViolet} \textit{Do the diagonals for the analog stick work correctly for you? [...]} } \end{tabular}
\\ \midrule
\textbf{Output} & Ranked list of passages & & Ranked list of responses \\ \hdashline
\begin{tabular}[c]{@{}l@{}}\textbf{Example} \end{tabular} & $p^+$: \textit{Thera-Derm Lotion is used as a moisturizer to treat or prevent dry, rough, scaly, [...]}  & & $r^+$: {\color{purple} \textit{In the "Others" tab, try [...]}}\\ \bottomrule
\end{tabular}
\end{table}

Given the differences between the domains, we verify empirically across three information-seeking datasets and 1.7M queries, the generalizability of five findings (\textbf{F1} to \textbf{F5}) from the passage and document retrieval literature related to state-of-the-art sparse and dense retrieval models. We are motivated in our selection of these five findings by their impact in prior works (cf. \S\ref{sec:related}). Our results show that four out of five previous findings do indeed generalize to our domain:
\begin{itemize}
    \item[\textbf{F1}] \xmark{}\footnote{{\color{red}\xmark{}} indicates that the finding does not hold in our domain whereas \cmark{} indicates that it holds in our domain followed by the \textit{necessary condition or exception}.} Dialogue context (i.e. query) expansion outperforms a no-expansion baseline~\cite{abdul2004umass,lin2019simplest,yang2019critically,lin2021pretrained}.
    \item[\textbf{F2}] \cmark{} Response (i.e. document) expansion outperforms a no-expansion baseline \cite{nogueira2019doc2query,lin2021pretrained,lin2021proposed} \textit{if the expansion model is trained to generate the most recent context (last utterance\footnote{For example in Table~\ref{table:intro} the last utterance is $u^3$.} of the dialogue) instead of older context (all utterances).}
    \item[\textbf{F3}] \cmark{} Dense retrieval in the zero-shot\footnote{A zero-shot is a model that does not have access to target data, cf. Table~\ref{table:stages}.} setting underperforms sparse baselines \cite{ren2022thorough,thakur2021beir} \textit{except when it goes through intermediate training on large amounts of out-of-domain data.}
    \item[\textbf{F4}] \cmark{} Dense retrieval with access to target data\footnote{Target data is data from the same distribution, i.e. dataset, of the evaluation dataset.} outperforms sparse baselines \cite{gao2021unsupervised,karpukhin2020dense,ren2022thorough} \textit{if an intermediate training step on out-of-domain data is performed before the fine-tuning on target data.}
    \item[\textbf{F5}] \cmark{} Harder negative sampling techniques lead to effectiveness gains \cite{xiong2020approximate,zhan2021optimizing} \textit{if a denoising technique is used to reduce the number of false negative samples.}
\end{itemize}

Our results indicate that most findings translate to the domain of retrieval of responses for dialogues. A promising future direction is thus to start with successful models from other domains---for which there are more datasets and previous research---and study how to adapt and improve them for retrieval-based conversational search.

\section{Related Work}

\label{sec:related}
In this section we first discuss current research in retrieval-based systems for conversational search, followed by reviewing the major findings of (un)supervised sparse and dense retrieval in the domains of passage and document retrieval.

\subsection{Ranking and Retrieval of Responses for Dialogues}

Early neural models for response re-ranking were based on matching the representations of the concatenated dialogue context and the representation of a response in a single-turn manner with architectures such as CNN and LSTM~\cite{lowe2015ubuntu,kadlec2015improved}. More complex neural architectures matching each utterance with the response were also explored ~\cite{zhou2018multi,gu2019interactive,lin2020world}. Heavily pre-trained language models such as BERT were first shown to be effective by Nogueira and Cho~\cite{nogueira2019passage} for re-ranking. Such models quickly became a predominant approach for re-ranking in IR~\cite{lin2021pretrained} and were later shown to be effective for re-ranking responses in conversations~\cite{whang2019effective,penha2019curriculum}. 


In contrast, the first-stage retrieval of responses for a dialogue received relatively little attention~\cite{penha2020challenges}. Lan et al.~\cite{lan2021exploring} and Tao et al.~\cite{tao2021building} showed that BERT-based dense retrieval models outperform BM25 for first-stage retrieval of responses for dialogues. A limitation of their work is that strong sparse retrieval baselines that have shown to be effective in other retrieval tasks, e.g. BM25 with \textit{dialogue context expansion}~\cite{nogueira2019doc2query} or BM25 with \textit{response expansion}~\cite{yang2019critically}, were not employed for dense retrieval. We do such comparisons here and test a total of five major findings that have been not been evaluated before by previous literature on the first-stage retrieval of responses for dialogues.






\subsection{Dense and Sparse Models for Passage and Document Retrieval}

\subsubsection{Context for F1}
Retrieval models can be categorized into two dimensions: supervised vs. unsupervised and dense vs. sparse representations~\cite{lin2021proposed}. An \emph{unsupervised} sparse representation model such as BM25~\cite{robertson1994some} represents each document and query with a sparse vector with the dimension of the collection's vocabulary, having many zero weights due to non-occurring terms. Since the weights of each term are entirely based on term statistics they are considered unsupervised methods. Such approaches are prone to the vocabulary mismatch problem~\cite{furnas1987vocabulary}, as semantic matches are not considered. A way to address such a problem is by using query expansion methods. RM3~\cite{abdul2004umass} is a competitive~\cite{yang2019critically} query expansion technique that uses pseudo-relevance feedback to add new terms to the queries followed by another final retrieval step using the modified query. 


\subsubsection{Context for F2}
A \emph{supervised} sparse retrieval model can take advantage of the effectiveness of transformer-based language models by changing the terms' weights from collection statistics to something that is learned. Document expansion with a learned model can be considered a learned sparse retrieval approach~\cite{lin2021proposed}. The core idea is to create pseudo documents that have expanded terms and use them instead when doing retrieval. Doc2query~\cite{nogueira2019doc2query} is a strong supervised sparse retrieval baseline that uses a language model to predict queries that might be issued to find a document. The predictions of this model are used to create the augmented pseudo documents.




\subsubsection{Context for F3 and F4}
Supervised dense retrieval models\footnote{A distinction can also be made of cross-encoders and bi-encoders, where the first encode the query and document jointly as opposed to separately~\cite{thakur2020augmented}. Cross-encoders are applied in a re-ranking step due to their inefficiency and thus are not our focus.}, such as ANCE~\cite{xiong2020approximate} and coCodenser~\cite{gao2021unsupervised}, represent query and documents in a small fixed-length space, for example of 768 dimensions. Dense retrieval models without access to target data for training---known as the \emph{zero-shot scenario}---have underperformed sparse methods (\textbf{F3}). For example, the \texttt{BEIR} benchmark~\cite{thakur2021beir} showed that BM25 was superior to dense retrieval from 9--18 (depending on the model) out of the 18 datasets in the zero-shot scenario. In contrast, when having access to enough supervision from target data, dense retrieval models have shown to consistently outperform strong sparse baselines~\cite{gao2021unsupervised,karpukhin2020dense,ren2022thorough} (\textbf{F4}).

\subsubsection{Context for F5}

In order to train neural ranking models, a small set of negative (i.e. non-relevant) candidates are necessary as it is prohibitively expensive to use every other document in the collection as negative sample for a query. A limitation of randomly selecting negative samples is that they might be too easy for the ranking model to discriminate from relevant ones, while for negative documents that are harder the model might still struggle. For this reason hard negative sampling has been shown to perform better than random sampling for passage and document retrieval~\cite{xiong2020approximate,robinson2020contrastive,zhan2021optimizing}.

\section{First-stage Retrieval for Dialogues} \label{section:method}


In this section we first describe the problem of first-stage retrieval of responses, followed by the findings we want to replicate from sparse and dense approaches.

\subsubsection{Problem Definition}
The task of first-stage retrieval of responses for dialogues, concerns retrieving the best response out of the entire collection given the dialogue context. Formally, let $\set{D}=\{(\set{U}_i, \set{R}_i, \set{Y}_i)\}_{i=1}^{M}$ be a data set consisting of $M$ triplets: dialogue context, response candidates and response relevance labels. The dialogue context $\set{U}_i$ is composed of the previous utterances $\{u^1, u^2, ... , u^{\tau}\}$ at the turn $\tau$ of the dialogue. The candidate responses $\set{R}_i = \{r^1, r^2, ..., r^n\}$ are either ground-truth responses $r^+$ or negative sampled candidates $r^-$, indicated by the relevance labels $\set{Y}_i = \{y^1, y^2, ..., y^n\}$. In previous work, the number of candidates is limited, typically $n = 10$~\cite{penha2020challenges}. The findings we replicate here come from passage and document retrieval tasks where there is no limit to the number of documents or passages that have to be retrieved. Thus, in all of our first-stage retrieval task experiments $n$ is set to the size of the entire collection of responses in the corpus. The number of ground-truth responses is one, the observed response in the conversational data. The task is then to learn a ranking function $f(.)$ that is able to generate a ranked list from the entire corpus of responses $\set{R}_i$ based on their predicted relevance scores $f(\set{U},r)$.

\subsubsection{F1: Unsupervised Sparse Retrieval}

We rely on classic retrieval methods, for which the most commonly used baseline is \bm{}. One of the limitations of sparse retrieval is the vocabulary mismatch problem. Expansion techniques are able to overcome this problem by appending new words to the dialogue contexts and responses. For this reason, we here translate a query expansion technique to the dialogue domain and perform \emph{dialogue context expansion} with RM3~\cite{abdul2004umass}, a competitive unsupervised method that assumes that the top-ranked responses by the sparse retrieval model are relevant. From these pseudo-relevant responses, words are selected and an expanded dialogue context is created and subsequently employed by the sparse retrieval method to rank the final list of responses. \textbf{The effectiveness of RM3 in the domain of dialogues is the first finding that we validate}.

\subsubsection{F2: Learned Sparse Retrieval} Alternatively, we can expand the responses in the collection with a learned method. To do so we ``translate'' doc2query~\cite{nogueira2019doc2query} into our domain, yielding \resptocontext{}. Formally, we fine-tune a generative transformer model $G$ for the task of generating the dialogue context $\set{U}_i$ from the ground-truth response $r^+_i$. This model is then used to generate expansions for all responses in the collection, $r^i=concat(r^i, G(r^i))$. These expansions are appended to the responses and the collection is indexed again---the sparse retrieval method itself is not modified, i.e. we continue using \bm{}. This approach (which we coin \resptocontext{}) leads to two improvements: term re-weighting (adding terms that already exist in the document) and dealing with the vocabulary mismatch problem (adding new terms). \textbf{The effectiveness of doc2query in the domain of dialogues is the second finding that we validate}.

Unlike passage and document retrieval where the queries are smaller than the documents, for the retrieval of responses for dialogues the queries are longer than the documents\footnote{For example, while the \texttt{TREC-DL-2020} passage and document retrieval tasks the queries have between 5--6 terms on average and the passages and documents have over 50 and 1000 terms respectively, for the information-seeking dialogue datasets used here the dialogue contexts (queries) have between 70 and 474 terms on average depending on the dataset while the responses (documents) have between 11 and 71.}. This is a challenge for the generative model, since generating larger pieces of text is a more difficult problem than smaller ones as there is more room for errors. Motivated by this, we also explored a modified version of \resptocontext{} that aims to generate only the last utterance of the dialogue context: \resptocontextlu{}. This model is trained to generate $u^{\tau}$ from $r^+_i$, instead of trying to generate the whole utterance $\set{U}_i = \{u^1, u^2, ... , u^{\tau}\}$. The underlying premise is that the most important utterance from the dialogue is the last one, and if it is correctly generated by \resptocontextlu{}, the sparse retrieval method will be able to find the correct response from the collection.

\subsubsection{F3: Zero-shot Dense Retrieval}

We rely on methods that learn to represent the dialogue context and the responses separately in a dense embedding space. Responses are then ranked by their similarity to the dialogue context. We rely here on pre-trained language transformer models, such as BERT~\cite{devlin2018bert} and MPNet~\cite{song2020mpnet}, to obtain such representations of the dialogue context and response. This approach is generally referred to as a \emph{bi-encoder} model~\cite{lin2021pretrained} and is an effective family of models\footnote{See for example the top models in terms of effectiveness from the MSMarco benchmark leaderboards \url{https://microsoft.github.io/msmarco/}.}. A zero-shot model is one that is not trained on the target data. Target data is data from the same distribution, i.e. dataset, of the evaluation dataset.


One way of improving the representations of a heavily pre-trained language model for the zero-shot setting is to fine-tune it with intermediate data~\cite{reimers-2019-sentence-bert}. Such intermediate data contains triplets of query, relevant document, and negative document and can include multiple datasets. The advantage of adding this step before employing the representations of the language model is to reduce the gap between the pre-training and the downstream task at hand~\cite{pruksachatkun2020intermediate,peeters2020intermediate,poth2021pre}. In Table~\ref{table:stages} we clarify the relationship between pre-training, intermediate training and fine-tuning.

\vspace{-0.5cm}
\begin{table}[]
\scriptsize
\centering
\caption{The different training stages and data, their purposes, examples of datasets, and the type of dense model obtained after each stage.}
\label{table:stages}
\begin{tabular}{@{}lp{3cm}p{3.75cm}p{3.75cm}@{}}
\toprule
                 & \textbf{Pre-training data}                      & \textbf{Intermediate data}        & \textbf{Target data}                          \\ \midrule
\textbf{Purpose} & Learn general representations & Learn sentence representations for ranking  & Learn representations for target distribution \\ 
\textbf{Model is} & Zero-shot & Zero-shot  & Fine-tuned \\
\textbf{Example} & \texttt{Wikipedia}                      & \texttt{MSMarco}                           & \mantis{}                                     \\ \bottomrule
\end{tabular}
\end{table}

The intermediate training step learns to represent pieces of text (query and documents) by applying a mean pooling function over the transformer's final layer, which is then used to calculate the dot-product similarity. The loss function employs multiple negative texts from the same batch to learn the representations in a constrastive manner, also known as in-batch negative sampling. Such a procedure learns better text representations than a naive approach that uses the $[CLS]$ token representation of BERT~\cite{reimers-2019-sentence-bert,aghajanyan2021muppet}. 


The function $f(\set{U},r)$ is then $dot(\eta(concat(\set{U})),\eta(r))$, where $\eta$ is the representation obtained by applying the mean pooling function over the last layer of the transformer model, and $concat(\set{U}) = u^1 \; | \; \usep \; | \; u^2 \; | \;  \tsep \; | \; ... \; | \; u^{\tau} \;$, where $|$ indicates the concatenation operation. The utterances from the context $\set{U}$ are concatenated with special separator tokens $\usep$ and $\tsep$ indicating end of utterances and turns\footnote{The special tokens $\usep$ and $\tsep$ will not have any meaningful representation in the zero-shot setting, but they can be learned on the fine-tuning step.}. \textbf{The effectiveness of a zero-shot bi-encoder model in the domain of dialogues is the third finding we validate.}

\subsubsection{F4: Fine-tuned Dense Retrieval}

The standard procedure is to fine-tune dense models with target data that comes from the same dataset that the model will be evaluated. Since we do not have labeled negative responses, all the remaining responses in the dataset can be thought of as non-relevant to the dialogue context. Computing the probability of the correct response over all other responses in the dataset would give us $P(r \mid \set{U})=\frac{P(\set{U}, r)}{\sum_{k} P\left(\set{U}, r_{k}\right)}$. This computation is prohibitively expensive, and the standard procedure is to approximate it using a few negative samples. The \emph{negative sampling} task is then as follows: given the dialogue context $\set{U}$ find challenging responses $r^-$ that are non-relevant for $\set{U}$. Negative sampling can be seen as a retrieval task, where one can use a model to retrieve negatives by applying a retrieval function to the collection of responses using $\set{U}$ as the query.


With such a dataset at hand, we continue the training---after the intermediate step---in the same manner as done by the intermediate training step, with the following cross-entropy loss function\footnote{We refer to this loss as MultipleNegativesRankingLoss.} for a batch with size $B$:

    $\mathcal{J}(\mathbf{\set{U}}, \mathbf{r}, \theta) =-\frac{1}{B} \sum_{i=1}^{B}\left[f\left(\set{U}_{i}, r_{i}\right)-\log \sum_{j=1, j!=i}^{B} e^{f\left(\set{U}_{i}, r_{j}\right)}\right],$

where $f(\set{U},r)$ is the dot-product of the mean pooling of the last layer of the transformer model. \textbf{The effectiveness of a fine-tuned bi-encoder model in the domain of dialogues is the fourth finding we validate here.}


\subsubsection{F5: Hard Negative Sampling}
A limitation of random samples is that they might be too easy for the ranking model to discriminate from relevant ones, while for negative documents that are hard the model might still struggle. For this reason, another popular approach is to use a ranking model to retrieve negative documents using the given query with a classic retrieval technique such as BM25. This leads to finding negative documents that are closer to the query in the sparse representation space, and thus they are \emph{harder negatives}. Since dense retrieval models have been outperforming sparse retrieval in a number of cases with available training data, more complex negative sampling techniques making use of dense retrieval have also been proposed~\cite{xiong2020approximate,hofstatter2021efficiently}. \textbf{The effectiveness of hard negative sampling for a bi-encoder model in the domain of dialogues is the fifth finding we validate here.}

\section{Experimental Setup}
In order to compare the different sparse and dense approaches we consider three large-scale information-seeking conversation datasets\footnote{\msdialog{} is available at~\url{https://ciir.cs.umass.edu/downloads/msdialog/}; \mantis{} is available at~\url{https://guzpenha.github.io/MANtIS/}; \ubuntu{} is available at ~\url{https://github.com/dstc8-track2/NOESIS-II}.}: \textbf{\msdialog{}}~\cite{qu2018analyzing} contains 246K context-response pairs, built from 35.5K information seeking conversations from the Microsoft Answer community, a QA forum for several Microsoft products; \textbf{\mantis{}}~\cite{penha2019introducing} contains 1.3 million context-response pairs built from conversations of 14 Stack Exchange sites, such as \textit{askubuntu} and \textit{travel}; \textbf{\ubuntu{}}~\cite{Kummerfeld_2019} contains 184k context-response pairs of disentangled Ubuntu IRC dialogues. 

\subsubsection{Implementation Details}
For BM25 and BM25+RM3\footnote{We perform hyperparameter tuning using grid search on the number of expansion terms, number of
expansion documents, and weight.} we rely on the \texttt{pyserini} implementations~\cite{Lin_etal_SIGIR2021_Pyserini}. In order to train \resptocontext{} expansion methods we rely on the Huggingface \texttt{transformers} library~\cite{wolf2019huggingface}, using the \textit{t5-base} model. We fine-tune the {T5} model for 2 epochs, with a learning rate of 2e-5, weight decay of 0.01, and batch size of 5. When augmenting the responses with \resptocontext{} we follow docT5query~\cite{nogueira2019doc2query} and append three different context predictions, using sampling and keeping the top-10 highest probability vocabulary tokens.

For the zero-shot dense models, we rely on the \texttt{SentenceTransformers}~\cite{reimers-2019-sentence-bert} model releases. The library uses Hugginface's \texttt{transformers} for the pre-trained models such as BERT~\cite{devlin2018bert} and MPNet~\cite{song2020mpnet}. 
For the 
bi-encoder models, we use the pre-trained \textit{all-mpnet-base-v2} weights which were the most effective in our initial experiments, compared with other pre-trained models\footnote{The alternative models we considered are those listed in the model overview section at \url{https://www.sbert.net/docs/pretrained\_models.html}.}. When fine-tuning the dense retrieval models, we rely on the \textit{MultipleNegativesRankingLoss}, which accepts a number of hard negatives, and also uses the remaining in-batch random negatives to train the model. We use a total of 10 negative samples for dialogue context. We fine-tune the dense models for a total of 10k steps, and every 100 steps we evaluate the models on a re-ranking task that selects the relevant response out of 10 responses. We use the re-ranking validation MAP to select the best model from the whole training to use in evaluation. We use a batch size of 5, with 10\% of the training steps as warmup steps. The learning rate is 2e-5 and the weight decay is 0.01. We use \texttt{FAISS}~\cite{johnson2019billion} to perform the similarity search.


\subsubsection{Evaluation}
To evaluate the effectiveness of the retrieval systems we use $R@K$. We thus evaluate the models' capacity of finding the correct response out of the whole possible set of responses\footnote{The standard evaluation metric in conversation response ranking~\cite{yuan2019multi,gu2020speaker,tao2019multi} is recall at position $K$ with $n$ candidates $R_n@K$. Since we are focused on the first-stage retrieval we set $n$ to be the entire collection of answers}. We perform Students t-tests at the 0.95 confidence level with Bonferroni correction to compare statistical significance of methods. Comparisons are performed across the results for each dialogue context.

\section{Results}

In this section, we discuss our empirical results along with the five major findings from previous work (Section~\ref{sec:intro}) in turn. Table~\ref{table:main_table_results} contains the main results regarding \textbf{F1} to \textbf{F4}.  Table~\ref{table:denoising} contains the results for \textbf{F5}. 

\begin{table}[ht!]
\centering
\caption{Results for the generalizability of F1--F4. Bold values indicate the highest recall for each type of approach. Superscripts indicate statistically significant improvements using Students t-test with Bonferroni correction. \textit{$\dagger$=other methods from the same group$1$=best from unsupervised sparse retrieval ; $2$=best from supervised sparse retrieval; $3$=best from zero-shot dense retrieval.} For example, in F3 $^\dagger$ indicates that row (3d) improves over rows (3a--c), $^1$ indicates that it improves over row (1a) and $^2$ indicates it improves over row (2b).}
\label{table:main_table_results}
\begin{tabular}{@{}llllllll@{}}
\toprule
 &   & \multicolumn{2}{c}{\textbf{\mantis{}}} & \multicolumn{2}{c}{\textbf{\msdialog{}}} & \multicolumn{2}{c}{\textbf{\ubuntu{}}} \\ \midrule
 &   & \textbf{R@1} & \textbf{R@10} & \textbf{R@1} & \textbf{R@10} & \textbf{R@1} & \textbf{R@10} \\ \midrule
(0) & Random & 0.000 & 0.000 & 0.000 & 0.001 & 0.000 & 0.001 \\ \midrule
 \multicolumn{5}{l}{\textbf{Unsupervised sparse}}   & \multicolumn{3}{l}{\textbf{F1}}  \\ \midrule
(1a) & \bm{} & \textbf{0.133$^{\dagger}$} & \textbf{0.299$^{\dagger}$} & \textbf{0.064$^{\dagger}$} & \textbf{0.177$^{\dagger}$} & \textbf{0.027$^{\dagger}$} & \textbf{0.070$^{\dagger}$} \\
(1b) & \bm{} + \rmthree{} & 0.073 & 0.206 & 0.035 & 0.127 & 0.011 & 0.049 \\ \midrule
 \multicolumn{5}{l}{\textbf{Supervised sparse}} & \multicolumn{3}{l}{\textbf{F2}} \\ \midrule
(2a) & \bm{} + \resptocontext{} & 0.135 & 0.309 & 0.074 & \textbf{0.208} & 0.028 & 0.067 \\
(2b) & \bm{} + \resptocontextlu{} & \textbf{0.147$^{\dagger 1}$} & \textbf{0.325$^{\dagger 1}$} & \textbf{0.075$^{1}$} & 0.202$^{1}$ & \textbf{0.029$^{}$} & \textbf{0.076$^{}$}\\ \midrule
 \multicolumn{5}{l}{\textbf{Zero-shot dense}{ (Model$_{IntermediateData}$)}} & \multicolumn{3}{l}{\textbf{F3}} \\ \midrule
(3a) & ANCE$_{600K-{MSMarco}}$  & 0.048 & 0.111 & 0.050 & 0.124 & 0.010 & 0.028 \\
(3b) & TAS-B$_{400K-{MSMarco}}$  & 0.062 & 0.143 & 0.060 & 0.157 & 0.019 & 0.050 \\
(3c) & Bi-encoder$_{215M-mul}$  & 0.138 & 0.297 & 0.108 & 0.277 & 0.023 & 0.076 \\
(3d) & Bi-encoder$_{1.17B-mul}$ & \textbf{0.155$^{\dagger 1}$} & \textbf{0.341$^{\dagger 12}$} & \textbf{0.147$^{\dagger 12}$} & \textbf{0.339$^{\dagger 12}$} & \textbf{0.041$^{\dagger}$} & \textbf{0.097$^{\dagger 12}$} \\ \midrule
 \multicolumn{5}{l}{\textbf{Fine-tuned dense}{ (Model$_{NegativeSampler}$)}} &  \multicolumn{3}{l}{\textbf{F4}} \\ \midrule
(4a) & Bi-encoder$_{Random (0)}$  & \textbf{0.130} & \textbf{0.307} & \textbf{0.168$^{123}$} & \textbf{0.387$^{123}$} & \textbf{0.050$^{12}$} & \textbf{0.128$^{123}$} \\
\bottomrule
\end{tabular}
\end{table}


\subsubsection*{\textbf{F1 \xmark{} Query expansion via RM3 leads to improvements over not using query expansion}~\cite{abdul2004umass,lin2019simplest,yang2019critically,lin2021pretrained}.}
\bm{}+\rmthree{} (row 1b) does not improve over \bm{} (1a) on any of the three conversational datasets analyzed. We performed thorough hyperparameter fine-tuning and no combination of the \rmthree{} hyperparameters outperformed \bm{}. \textbf{This indicates that F1 does not hold for the task of response retrieval for dialogues.}

A manual analysis of the new terms appended to a sample of 60 dialogue contexts by one of the paper's authors revealed that only 18\% of them have at least one relevant term added based on our best judgment. Unlike web search where the query is often incomplete, under-specified, and ambiguous, in the information-seeking datasets employed here the dialogue context (query) is quite detailed and has more terms than the responses (documents). We hypothesize that because the dialogue contexts are already quite descriptive, the task of expansion is trickier in this domain and thus we observe many dialogues for which the added terms are noisy.



\subsubsection*{\textbf{F2 \cmark{} Document expansion via \resptocontext{} leads to improvements over no expansion}~\cite{nogueira2019doc2query,lin2021pretrained,lin2021proposed}.}
We find that a naive approach to response expansion improves marginally in two of the three datasets with \bm{}+\resptocontext{} (2a) outperforming \bm{} (1a). However, the proposed modification of predicting only the last utterance of the dialogue (\resptocontextlu{}) performs better than predicting the whole utterance, as shown by \bm{}+\resptocontextlu{}'s (2b) higher recall values. In the \mantis{} dataset the R@10 goes from 0.309 when using the model trained to predict the dialogue context to 0.325 when using the one trained to predict only the last utterance of the dialogue context.
\textbf{We thus find that F2 generalizes to response retrieval for dialogues, especially when predicting only the last utterance of the context\footnote{As future work, more sophisticated techniques can be used to determine which parts of the dialogue context should be predicted.}.}


\begin{table}[]
\small
\centering
\caption{Statistics of the augmentations for \resptocontext{} and \resptocontextlu{}. New words are the ones that did not exist in the document before.}
\label{table:resp2context_stats}
\begin{tabular}{@{}lrrr@{}}
\toprule
 & \textbf{\mantis{}} & \textbf{\msdialog{} }& \textbf{\ubuntu{}} \\ \midrule
Context avg length & 474.12 & 426.08 & 76.95 \\
Response avg length & 42.58 & 71.38 & 11.06 \\ \hdashline
Aug. avg length - \resptocontext{} & 494.23 & 596.99 & 202.3 \\
Aug. avg length - \resptocontextlu{} & 138.5 & 135.29 & 72.57 \\
\% new words  -  \resptocontext{} & 71\% & 69\% & 71\% \\
\% new words  -  \resptocontextlu{} & 59\% & 37\% & 63\% \\ \bottomrule
\end{tabular}
\end{table}

In order to understand what the response expansion methods are doing most---term re-weighting or adding novel terms---we present the percentage of novel terms added by both methods in Table~\ref{table:resp2context_stats}. The table shows that \resptocontextlu{} does more term re-weighting than adding new words when compared to \resptocontext{} (53\% and 70\% on average are new words respectively and thus 47\% vs 30\% are changing the weights by adding existing words), generating overall smaller augmentations (115.45 vs 431.17 on average respectively). 



\subsubsection*{\textbf{F3 \cmark{} Sparse retrieval outperforms zero-shot dense retrieval}~\cite{ren2022thorough,thakur2021beir}.}

Sparse retrieval models are more effective than the majority of zero-shot dense models, as shown by the comparison of rows (1a--b), and (2a--b) with rows (3a--c). However, a dense retrieval model that has gone through intermediate training on large and diverse datasets including dialogues is more effective than a strong sparse retrieval model, as we see by comparing row (3d) with row (2b) in Table~\ref{table:main_table_results}. 

For example, while the zero-shot dense retrieval models based only on the \texttt{MSMarco} dataset (3a--b) perform on average 35\% worse than the strong sparse baseline (2b) in terms of R@10 for the \msdialog{} dataset, the zero-shot model trained with 1.17B instances on diverse data (3d) is 68\% better than the sparse baseline (2b). When using a bigger amount of intermediate training data\footnote{For the full description of the intermediate data see \url{https://huggingface.co/sentence-transformers/all-mpnet-base-v2}.}, we see that the zero-shot dense retrieval model (3d) is able to outperform the sparse retrieval baseline by margins of 33\% of R@10 on average across datasets.

\textbf{We thus show that F3 only generalizes to response retrieval for dialogues if we do not employ a large set of diverse intermediate data.} As expected, the closer the intermediate training data distribution is to the evaluation data, the better the dense retrieval model performs. The results indicate that a good zero-shot retrieval model needs to go through intermediate training on a large set of training data coming from multiple datasets to generalize well to different domains and outperform strong sparse retrieval baselines.

\subsubsection*{\textbf{F4 \cmark{} Dense models with access to target training data outperform sparse models}~\cite{gao2021unsupervised,karpukhin2020dense,ren2022thorough}.}

First, we see that fine-tuning the dense retrieval model, which has gone through intermediate training already, with random sampling---row (4a) in Table~\ref{table:main_table_results}---achieves the best overall effectiveness in two of the three datasets. \textbf{This result shows that F4 generalizes to the task of response retrieval for dialogues when employing intermediate training\footnote{Our experiments show that when we do not employ the intermediate training step the fine-tuned dense model does not generalize well, with row (3d) performance dropping to 0.172, 0.308 and 0.063 R@10 for \mantis{}, 
\msdialog{} and \ubuntu{} respectively.}}. Having access to the target data as opposed to only the intermediate training data means that the representations learned by the model are closer to the true distribution of the data.

We hypothesize that fine-tuning the bi-encoder for \mantis{} (4a) is harmful because the intermediate data contains Stack Exchange responses. In this way, the set of dialogues of Stack Exchange that \mantis{} encompasses might be serving only to overfit the intermediate representations. As evidence for this hypothesis, we found that (I) the learning curves flatten quickly (as opposed to other datasets) and (II) fine-tuning another language model that does not have Stack Exchange data (\texttt{MSMarco}) in their fine-tuning, bi-encoder$_{bert-base}$ (3c), improves the effectiveness with statistical significance from 0.092 R@10 to 0.205 R@10. 

\subsubsection*{\textbf{F5 \cmark{} Hard negative sampling is better than random sampling for training dense retrieval models~\cite{xiong2020approximate,zhan2021optimizing}}.}

Surprisingly we find that naively using more effective models to select negative candidates is detrimental to the effectiveness of the dense retrieval model (see Hard negative sampling in Table~\ref{table:denoising}). We observe this phenomenon when using different language models, when switching intermediate training on or off for all datasets, and when using an alternative contrastive loss~\cite{hadsell2006dimensionality} that does not employ in-batch negative sampling\footnote{The results are not shown here due to space limitations}.


After testing for a number of hypotheses that might explain why harder negatives do not improve the effectiveness of the dense retrieval model, we found that false negative samples increase significantly when using better negative sampling methods. False negatives are responses that are potentially valid for the context. Such relevant responses lead to unlearning relevant matches between context and responses as they receive negative labels. See below an example of a false negative sample retrieved by the bi-encoder model (row 3d of Table~\ref{table:main_table_results}):
    
    \vspace{3mm} 
     \fbox{
     \centering
     \begin{minipage}{30em}
     \scriptsize
     \centering
    \begin{itemize}[leftmargin=-0.1cm] 
        \item[] \textbf{Dialogue context ($\set{U}$)}: {\color{BlueViolet} hey... how long until dapper comes out?} $\usep$ {\color{purple}14 days} [...] $\usep$ {\color{BlueViolet}i thought it was coming out tonight}
        \item[] \textbf{Correct response ($r^{+}$)}: {\color{purple}just kidding couple hours}
        \item[] \textbf{False negative sample ($r^{-}$)}: {\color{purple}there is a possibility dapper will be delayed [...] meanwhile, dapper discussions should occur in ubuntu+1}
    \end{itemize}
    \end{minipage}}
    \vspace{3mm} 

Denoising techniques try to solve this problem by reducing the number of false negatives. We employ a simple approach that instead of using the top-ranked responses as negative responses, we use the bottom responses of the top-ranked responses as negatives\footnote{For example, if we retrieve $k=100$ responses, instead of using responses from top positions 1--10, we use responses 91--100 from the bottom of the list.}. This decreases the chances of obtaining false positives and if $k << |\set{D}|$ we will not obtain random samples. Our experiments in Table~\ref{table:denoising} reveal that this denoising technique, row (3b), increases the effectiveness for harder negative samples, beating all models from Table~\ref{table:main_table_results} for two of the three datasets. \textbf{The results indicate that F5 generalizes to the task of response retrieval for dialogues only when employing a denoising technique.}

\begin{table}[]
\caption{Results for the generalizability of F5---with and without a denoising strategy for hard negative sampling. Superscripts indicate statistically significant improvements using Students t-test with Bonferroni correction .\textit{ $^{\dagger}$=significance against the random sampling baseline, $^{\ddagger}$=significance against hard negative sampling without denoising.}}
\label{table:denoising}
\centering
\begin{tabular}{@{}llllllll@{}}
\toprule
 &   & \multicolumn{2}{c}{\textbf{\mantis{}}} & \multicolumn{2}{c}{\textbf{\msdialog{}}} & \multicolumn{2}{c}{\textbf{\ubuntu{}}} \\ \midrule
 &   & & \textbf{R@10} &  & \textbf{R@10} &  & \textbf{R@10} \\ \midrule
 & \textbf{Baseline}  &  &  &  &  &   \\ \midrule
& (1) Bi-encoder$_{Random}$  & &  0.307 &  & 0.387 &  & \textbf{0.128} \\
 \midrule
 & \textbf{Hard negative sampling}  &  &  &  &  &   \\ \midrule
 & (2a) Bi-encoder$_{\bm{}}$  &  & 0.271 &  & 0.316 & & 0.087 \\
 & (2b) Bi-encoder$_{Bi-encoder}$ & & 0.146 & & 0.306 & & 0.051 \\ \midrule
  & \textbf{Denoised hard negative sampling}  &  &  &  &  &   \\ \midrule
 & (3a) Bi-encoder$_{\bm{}}$  &  & 0.257 & & 0.358$^{\ddagger}$ & & 0.121$^{\ddagger}$ \\
 & (3b) Bi-encoder$_{Bi-encoder}$ & & \textbf{0.316}$^{\dagger \ddagger}$ & & \textbf{0.397}$^{\dagger \ddagger}$ & & 0.107$^{\ddagger}$ \\ \bottomrule
\end{tabular}
\end{table}
\vspace{-0.5cm}

\section{Conclusion}
In this work, we tested if the knowledge obtained in dense and sparse retrieval from experiments on the tasks of passage and document retrieval generalizes to the first-stage retrieval of responses for dialogues. Our replicability study reveals that while most findings do generalize to our domain, a simple translation of the models is not always successful. A careful analysis of the domain in question might reveal better ways to adapt techniques.

As future work, we believe an important direction is to evaluate learned sparse methods that do weighting and expansion for both the queries and documents~\cite{formal2021splade}---while \resptocontext{} is able to both change the weights of the terms in the response (by repeating existing terms) and expand terms (by adding novel terms), it is not able to do weighting and expansion for the dialogue contexts.





\small
\subsubsection*{Acknowledgements}
This research has been supported by NWO projects SearchX (639.022.722) and NWO Aspasia (015.013.027).
%
%
%
\bibliographystyle{splncs04}
\bibliography{references}

\end{document}